\documentclass[10pt,twoside]{Lowx2011}
\usepackage{epsf,amsmath}
\usepackage{graphicx}

\def\beq{\begin{equation}}
\def\eeq{\end{equation}}
\def\bea{\begin{eqnarray}}
\def\eea{\end{eqnarray}}
\def\eq#1{{Eq.~(\ref{#1})}}
\def\fig#1{{Fig.~\ref{#1}}}

\newcommand{\Lb}{\left(}
\newcommand{\Rb}{\right)}

\begin{document}

\title{Gluon-decay cascade in AA collisions at the LHC and saturation}

\author{Amir H. Rezaeian \\ \\
 {\sl Departamento de F\'\i sica, Universidad T\'ecnica Federico Santa Mar\'\i a}\\
 {\sl  Avda. Espa\~na 1680, Casilla 110-V, Valparaiso, Chile } \\ \\
E-mail: Amir.Rezaeian@usm.cl}

\maketitle

\begin{abstract}
\noindent 
We extract the energy-dependence of the gluon-jet decay cascade from
$e^+e^-$ annihilation data and show that this energy-dependence has
power-low behavior for the average transverse momentum of jet bigger
than 1 GeV. We show that the observed different power-law
energy-dependence of hadron multiplicity in AA compared to pp
collisions at the LHC is due to the enhanced gluon-decay effect before hadronization. This
effect is more important for AA collisions where the saturation scale
is larger than 1 GeV.  Here we confront our published predictions in
arXiv:1102.2385 for the rapidity distribution of the charged hadron
production with the recently released data from the CMS collaboration
in AA collisions and show that our predictions is in perfect agreement
with the data.
 
\end{abstract}



\markboth{\large \sl \hspace*{0.25cm}A. H. Rezaeian 
\hspace*{0.25cm} Low-$x$ Meeting 2011} {\large \sl \hspace*{0.25cm} 
GLUON-DECAY CASCADE IN AA COLLISIONS ...}

\section{Introduction}
One of the most unexpected new feature of the LHC data has been the
very different power-law energy behavior of the charged hadron
multiplicities in AA compared to pp collisions \cite{Apb1,CMS}. There
have been several approaches to address the above problem
\cite{me1,JA,lap,al}. This is closely related to the open problem of
entropy production at the early stage of heavy-ion collisions (Mini
Bangs).

In the following, we provide a simple explanation of this observation and confront our published
predictions \cite{me1} with the recently released data from the LHC
\cite{CMS} for the rapidity distribution of the charged hadron production.  
 
\section{Main formalism and predictions}

\begin{figure}[t]
       \includegraphics[width=10 cm] {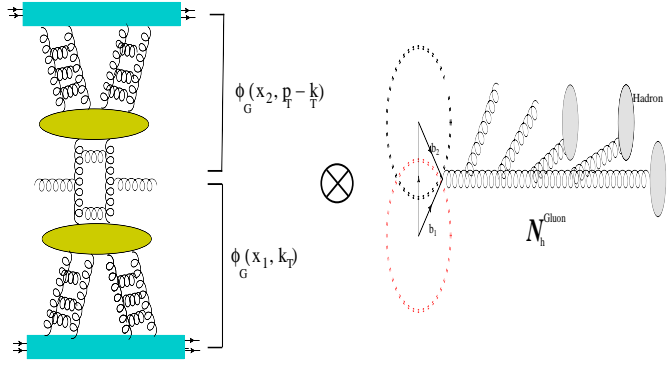}
\caption{ The corresponding diagram to the factorization given in \eq{I1}. In the $k_T$ factorization, the gluon production can be calculated as a convolution of  two unintegrated gluon density $\phi_G$ of the projectile and target \cite{KTINC}.   }
\label{f1}
\end{figure}

In the Color Glass Condensate (CGC)/saturation based approaches (for a
recent review see Ref.~\cite{review}), the hadron production may be
divided in two stages: production of gluons and subsequently the decay
of gluon-jet (or mini-jet) into hadrons.  Therefore, the multiplicity
of the produced hadrons at pseudo-rapidity $\eta$ can be calculated as
a convolution of these two stages (see \fig{f1}):
\bea \label{I1}
\frac{d N_h}{d \eta \,d^2 p_T}\,\,&\propto &\,\, \frac{d N^{Gluon}}{d y \,d^2 p_T} \otimes N^{Gluon}_h( E_{jet}), \\
\frac{d N_h}{d \eta }&\propto & \, \sigma_s Q^2_s \times N^{Gluon}_h\Lb Q_s\Rb,   \label{I11}\
\eea
where $\frac{d N^{Gluon}}{d yd^2 p_T}$ gives the gluon jet production
yield at rapidity $y$ (in pp or AA collisions) computable in the
$k_T$ factorization scheme \cite{me1,me2,KTINC}. The second term $N^{Gluon}_h$ is the
average multiplicity of hadrons in the gluon jet with a jet energy
$E_{jet}$ computable in the Modified Leading Logarithmic Approximation
(MLLA) scheme \cite{me2,ML}. The symbol $\otimes$ indicates a convolution, that is,
integrals over variables with possible weight factors included \cite{me1}.

The greatest simplicity of the CGC approach is the fact that all
complexity of infinite-body problem at very high energy (or small
$x$) shall be reduced into a one scale problem, with a hard saturation
scale $Q_s$ as the only dimensional relevant scale at which nonlinear
gluons recombination effects start to become important. In this
framework, the secondary hadrons are originated from the decay of
gluon mini-jets with the transverse momentum equal to the saturation
scale $Q_s$ \cite{me1,me2,me3}. Then \eq{I11} up to a possible
logarithmic correction, can be readily obtained by a dimensionality
argument where $\sigma_s$ is the effective area of interaction.

Notice that the $k_T$ factorization has been proven at the leading
$\log(1/x)$ approximation for scatterings of a dilute system on a
dense one (such as proton-nucleus collisions) and includes BFKL type
gluon emissions (with gluon fusion effects) between the projectile and
target, and also gluon radiations from the produced gluons
\cite{KTINC}. However, the other contribution of the gluon decay,
before hadronization, stems from the kinematic region outside the BFKL
emission regime where both emitted gluons are collinear to the emitter
\cite{me1}. In this kinematic region, the angle between the gluon
(quark) and the decay gluon is small and the main contribution of
gluon-decay has an opposite angular ordering to the BFKL type gluon
emissions. The later kinematic region has been traditionally studied
within a different resummation scheme, the so-called MLLA where it
contains systematically next-to-leading logarithmic corrections and
incorporates single and double-logarithmic effects in the development
of parton cascades \cite{ML}. Notice that the gluon decay effect
incorporated within these two resummation schemes have different
kinematics \cite{me1,KTINC,ML}.

\begin{figure}[t]
\centering
       \includegraphics[scale=0.47] {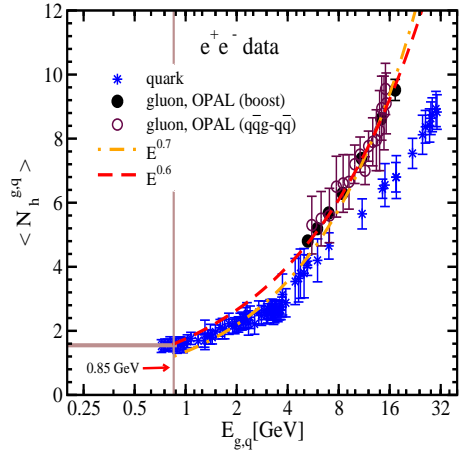}

\caption{The mean charged hadron multiplicity of unbiased gluon $N^g_h$ and quark $N^q_h$ jets in $e^+e^-$ annihilation, as a function of the jet energy. The plot is taken from \cite{me1}. }
\label{f2}
\end{figure}

\begin{figure}[t]
             \includegraphics[scale=0.5]{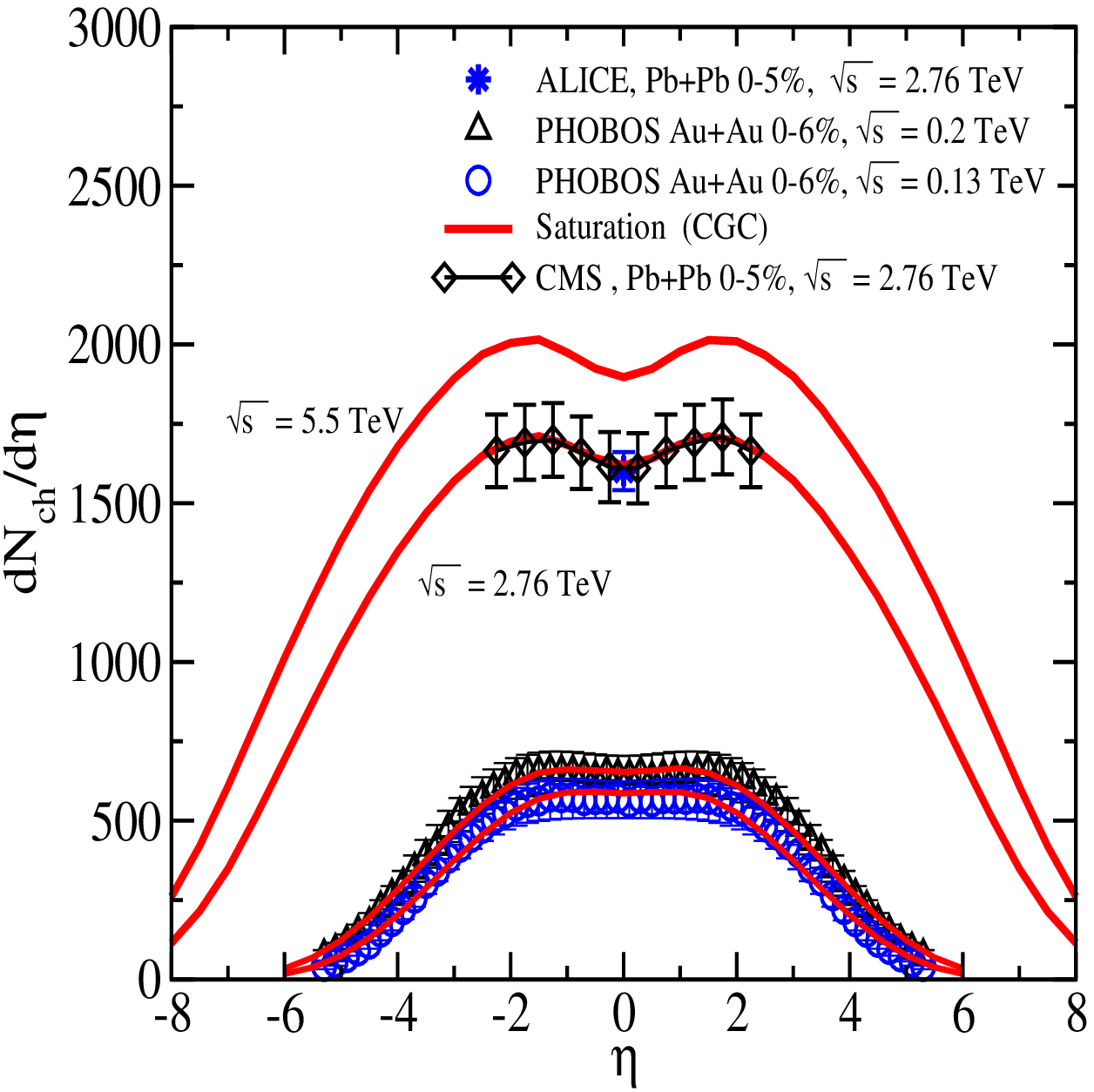}
       \includegraphics[scale=0.5] {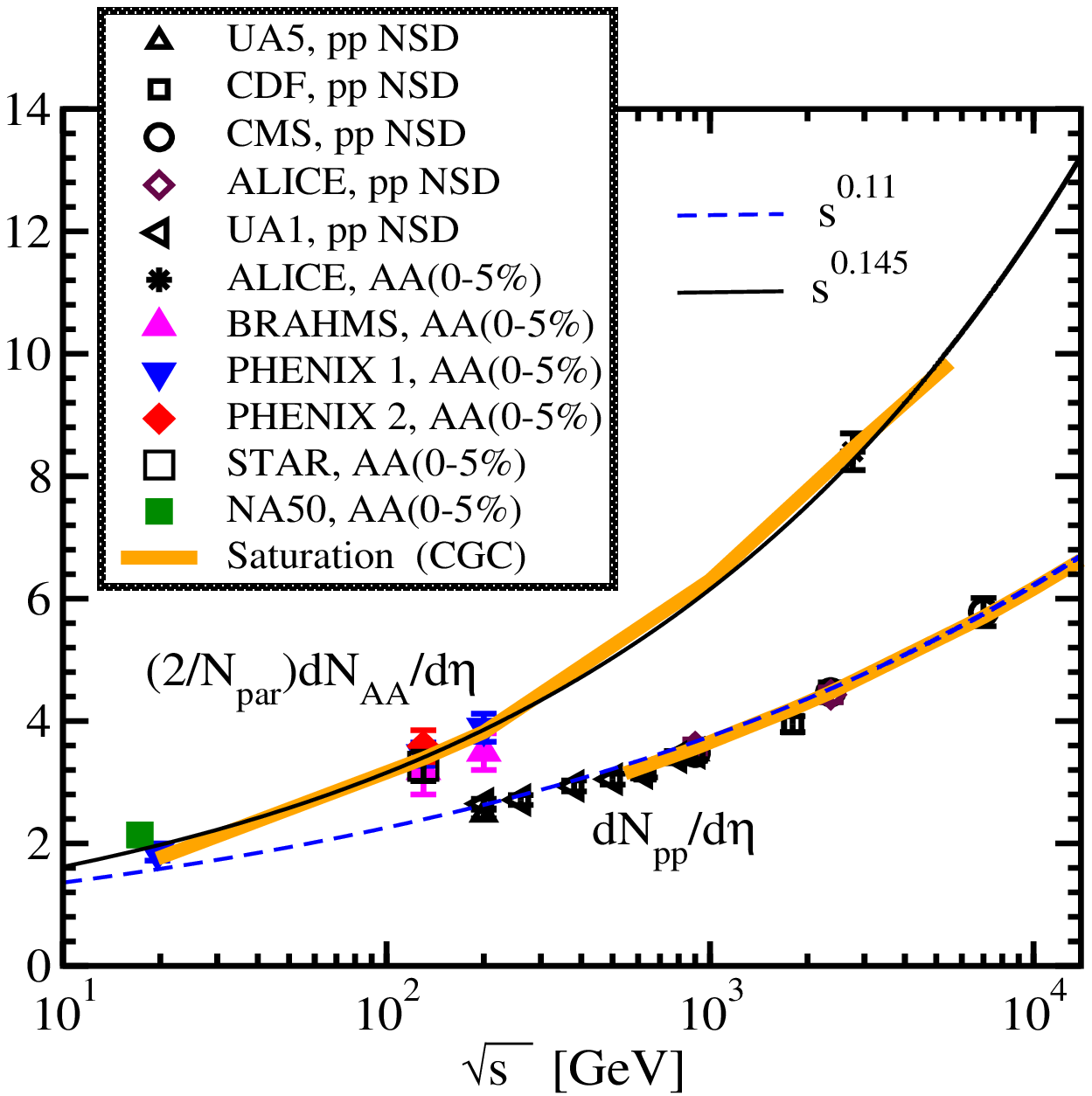}
\caption{Right: The energy behavior of charged particle pseudo-rapidity per participant pair for central AA and non-singlet diffractive pp collisions. The right panel plot is taken from Ref.~\cite{me1}. 
Left: Pseudo-rapidity distribution of charged particles
             produced at RHIC
             and the LHC. The experimental data from the LHC are from \cite{CMS,Apb1}. }
\label{f3}
\end{figure}
It is well-known that the gluon decay probability can be factorized
from the rest of cross-section in $e^+e^-\to q\bar q g$ reaction
\cite{ML}. This is in a sense, the essence of the factorization given in
\eq{I1}. One may extract information about the gluon-decay
stage in the MLLA region from gluon jet data in $e^+e^-$ collisions.
In order to obtain the energy dependence of the function $N^{Gluon}_h$, we use
directly experimental data for $N^{Gluon}_h \Lb E_{jet}\Rb$ in $e^+ e^-$ annihilation \cite{me1}. One can see from
\fig{f2} that $N^{Gluon}_h $ is constant at about $E_{jet} < 1$ GeV and
it grows as a power of $E_{jet}$ at higher energies.  From the
available $e^+e^-$ collisions data shown in \fig{f2}, we found that the energy-dependence
of the mean charged particle multiplicity of gluon-jet can be approximately described by 
\beq \label{eee}
 \langle N^{Gluon}_h\rangle \,\propto \,E^{\delta}_{jet},~~~\text{with}~~\delta=0.6\div 0.7 ~~~\text{for}~~~ E_{jet}\geq 0.85 \div 1~ \text{GeV}. 
\eeq
It is essential to stress again that such behavior also follows from
the theoretical estimates in the 3NLO pQCD in the MLLA scheme
\cite{ML}. Now, using Eqs.~(\ref{I11},\ref{eee}) and assuming the
typical energy of the gluon jet to be of the order of average saturation
scale, we obtain,
\begin{figure}[t]
              \includegraphics[scale=0.47]{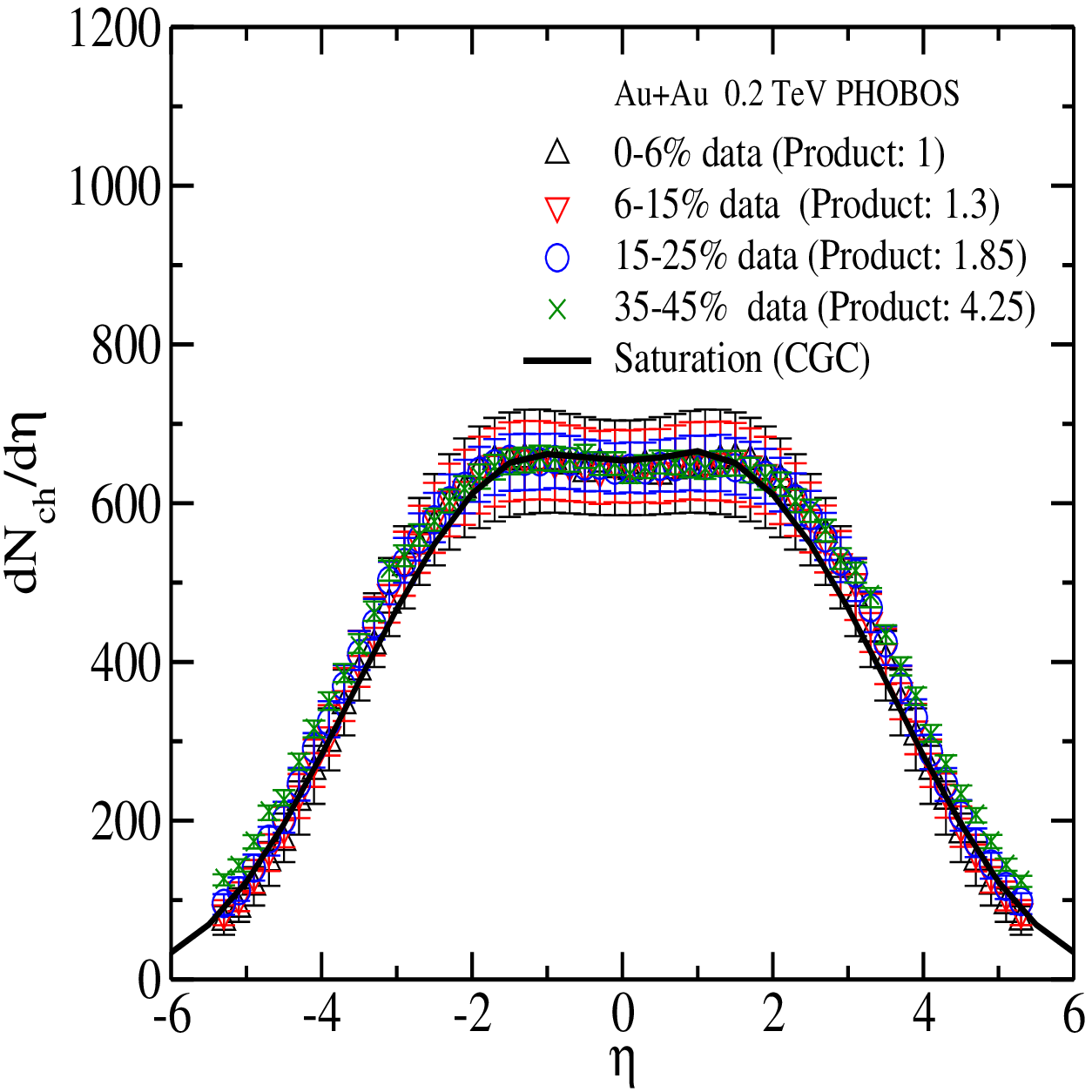}
               \includegraphics[scale=0.47] {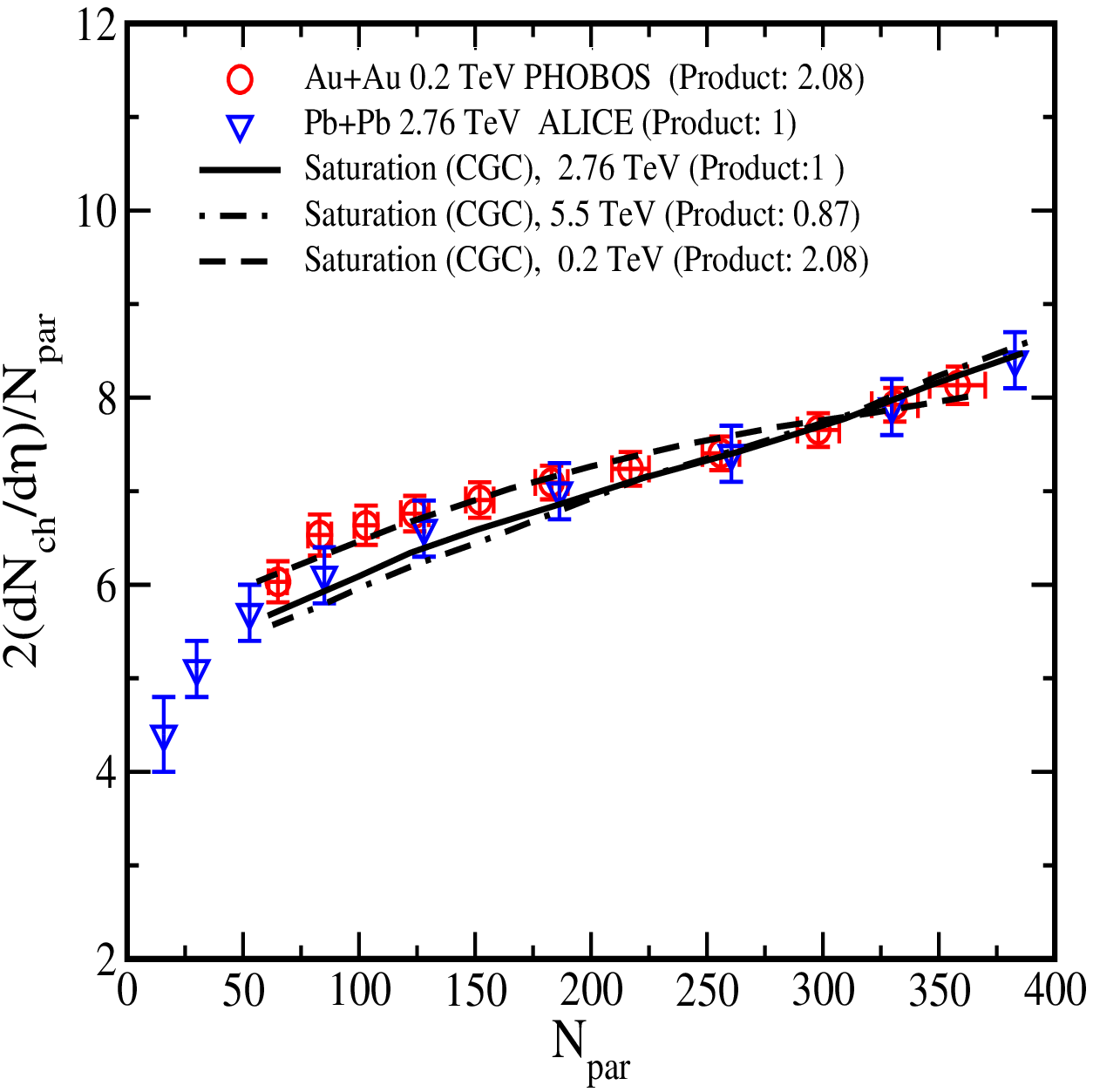}
\caption{Right: The scaled pseudo-rapidity density as a function of number of participant at midrapidity for AA collisions at various energies.  Left: The pseudo-rapidity distribution at RHIC $0.2$ TeV at different centralities.   
 The plots are taken from Ref.~\cite{me1}.}
\label{f4}
\end{figure}
\bea 
  \frac{dN_h}{d \eta}\Lb p p \Rb\,\,&\propto&
  \,\,Q^2_s\,\,\propto\,\,s^{\lambda/2} \,\,=\,\,s^{0.11}\,,
  \label{N21} \\ \frac{dN_h}{d \eta}\Lb A A \Rb\,\,&\propto& \,\,Q^2_s
  \times \Lb E_{jet}\,\propto \,Q_s\Rb^{0.65}\,\,\propto
  \,\,s^{\lambda/2+ 0.65\times\lambda/4}\,\,=\,\,s^{0.145},\label{N22}
  \eea 
where $s$ is the center-of-mass energy squared per nucleon pair and
$\lambda$ is free parameter to be fixed with other experiments like
DIS at HERA. In the above we assumed that the saturation scale for
pp collisions is $Q_s<1$ GeV and for AA collisions we have $Q_s>1$
GeV.  We take for the parameter
$\delta$, the average value $\overline \delta =0.65$ from \eq{eee}. In
\eq{N21}, the average value of $\lambda\approx 0.22$ in the the
effective saturation scale for pp collisions can be obtained from
$k_T$ factorization results given in Ref.~\cite{me2}. Then, the
power-law behavior given in \eq{N22} for AA collisions comes
naturally without any extra freedom. The simple
Eqs.~(\ref{N21},\ref{N22}) are enough to describe energy-dependence of
all existing data on hadron multiplicity from RHIC to the LHC, see
\fig{f3} (right). Indeed the full calculation \cite{me1} (shown by orange line in
\fig{f3}) based on the factorization shown in \fig{f1}, agrees with the
above argument based on the dimensionality. It is seen from Eqs.~(\ref{eee},\ref{N22}) that the
$k_T$ factorization accounts for the most energy-dependence of the
hadron multiplicity $s^{0.11}$, while the MLLA gluon-decay cascade (extracted from $e^+e^-$ annihilation data)
brings only about $s^{0.036}$ extra effect.  Notice that in the above
we assumed that the atomic number or $A$ (size of nuclei) dependence of
the saturation scale is factorizable from energy. This
assumption can be simply justified from the observation that the
centrality dependence of the multiplicity at the LHC is very similar
to RHIC up to an overall scale, see \fig{f4}.  This means that the energy
and centrality or $A$ dependence of multiplicity (and saturation scale) can be scaled out, in
agreement with our assumption. Therefore, data seems to role out any
extra energy dependence to be generated due to the geometry and the
centrality of collisions. In \fig{f3}, we compare our published prediction \cite{me1} for
the rapidity distribution of charged hadrons coming from the full
calculation with the recent CMS data \cite{CMS}. It is seen
in \fig{f3} that our predictions is in excellent agreement with data.
Our prediction for $dN_{AA}/d\eta$ at midrapidity for $0-5\%$ Pb+Pb
collisions at $5.5$ TeV is $1897\pm 133$ \cite{me1}.  We also show in  
\fig{f3} our prediction for the rapidity distribution of charged hadrons multiplicity for Pb+Pb collisions at $5.5$ TeV \cite{me1}.

\begin{figure}[t]
\includegraphics[scale=0.47] {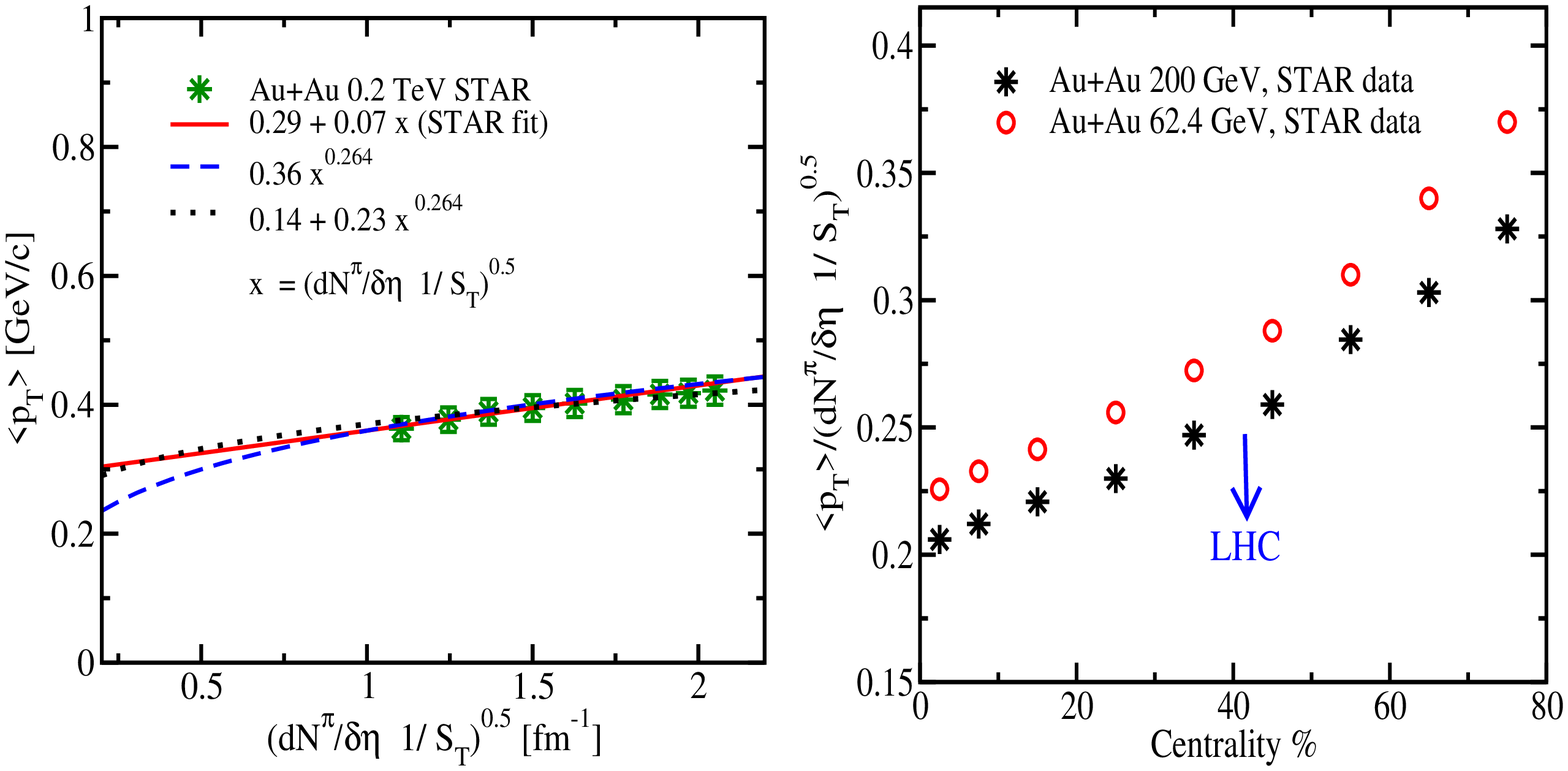}
\caption{Right: The ratio $\langle p_T
\rangle/\sqrt{(dN^{\pi}/d\eta)/S_T}$ at various centralities at RHIC. 
The data was constructed from three experimental measurements, the average transverse
momentum $\langle p_T \rangle$ of $\pi^-$, $dN^{\pi}/d\eta$ and $S_T$ \cite{star-f}. Left: Average transverse momenta as a
function of $\sqrt{(dN^{\pi}/d\eta)/\sigma_s}$ at RHIC. The experimental data are from \cite{star-f}. }
\label{f5}
\end{figure}

In \fig{f4} (right), we show $(2/N_{par})(dN_{AA}/d\eta)$ at
midrapidity (where $N_{par}$ is the number of participant for a given
centrality) has the scaling property at different energies. Notice that one can already
observe similar scaling property at RHIC, namely
$dN_{AA}/d\eta$ at fixed energy but different centralities falls into
a single curve upto a normalization factor, see \fig{f4} (left). Both
scaling properties shown in \fig{f4} can be easily understood within
the CGC picture and follows from simple \eq{I11}. We expect that the
centrality-scaling for $dN_{AA}/d\eta$ at a fixed energy will be also
valid at the LHC.  We predict that $(2/N_{par})(dN_{AA}/d\eta)$ for
$5.5$ TeV AA collisions at midrapidity to be about $\frac{1}{0.87\pm
0.06}$ times bigger than the corresponding one at $2.76$ TeV.

In \fig{f5} (right) we show the experimental data from the STAR
collaboration \cite{star-f} for $\langle p_T
\rangle/\sqrt{(dN^{\pi}/d\eta)/S_T}$ as a function of centrality where 
$\langle p_T \rangle$ is the average transverse momentum and $S_T$ is
the overlap area between the colliding nuclei in the transverse
plane. In our approach, we have $\langle p_T
\rangle/\sqrt{(dN/d\eta)/S_T}\sim
\frac{1}{n\sqrt{n}}$  where $n\sim N^{Gluon}_{h}$ for
$Q_s > 0.85 \div 1$ GeV corresponding to the excess of charged hadron
production in the presence of jet-decay effects. 
 It is seen from \fig{f5} that the ratio $\langle p_T
\rangle/\sqrt{(dN/d\eta)/S_T}$ at RHIC  Au+Au collisions
decreases for more central collisions and higher energies in accordance with our model and in contrast to the KLN type
approach \cite{KLN,lap}. We expect that the ratio $\langle p_T
\rangle/\sqrt{(dN/d\eta)/S_T}$ will be further suppressed at the LHC compared to RHIC, see \fig{f5}.  
Moreover, in the KLN type approaches \cite{KLN} we have $\langle p_T \rangle\sim x$ where
$x=\sqrt{(dN/d\eta)/S_T}$ while in our approach we have $\langle p_T
\rangle\sim x^{0.264}$ for the case that the saturation scale is $Q_s>0.85\div 1$ GeV. 
In \fig{f5} (left) we show
the average transverse momenta as a function of
$\sqrt{(dN/d\eta)/S_T}$.  The STAR collaboration \cite{star-f} has
found that the experimental data for the charged pion at different
centralities can be described by $\langle p_T
\rangle\approx p_0 + 0.07x$ where the constant $p_0=0.29$ GeV was obtained from a fit and may be interpreted as primordial transverse momentum. In \fig{f5}, we also show that a
fit driven by our approach prefers a smaller primordial transverse
momentum of about pion mass $p_0\sim 0.14~\text{GeV}$ (or even
$p_0\sim 0$) and it reasonably describes the same data. Notice that at small multiplicity for very peripheral
collisions at RHIC energy  entire saturation formulation is questionable.

To conclude: we extracted the energy-dependence of the gluon-jet decay
cascade from $e^+e^-$ annihilation data. We showed that the
energy-dependence of about $s^{0.036}$ due to the gluon-decay cascade
(when the average transverse momentum of the jet becomes about or
bigger than $1$ GeV) is exactly what explains the different power-law
energy-dependence of hadron multiplicity in AA compared to pp
collisions at the LHC. This effect is more important for AA collisions
where the saturation scale is larger and gives rise to an extra
contribution about $20-25\%$ to the multiplicity in AA collisions at
the LHC. On the theory side, this emphasizes on the importance of
higher order corrections beyond the leading log approximation in the $k_T$
factorization, and on general, this also gives rise to the outstanding
problem that how the fragmentation processes can be accommodated
within the CGC/saturation framework.

{\bf Acknowledgments.} 
The author would like to thank Genya Levin for the collaboration on this topic. The 
author is grateful to Yuri Kovchegov for insightful discussion. This work is
supported in part by Fondecyt grants 1110781.

\end{document}